%% file: ms.tex
\documentclass[twocolumn,tighten,10pt]{aastex631}

\usepackage{amsmath}

\newcommand{\PMO}{Purple Mountain Observatory, Chinese Academy of Sciences, Nanjing 210008, China}
\newcommand{\USTC}{School of Astronomy and Space Sciences, University of Science and Technology of China, Hefei 230026, China}
\newcommand{\YNU}{South-Western Institute for Astronomy Research, Yunnan University, Kunming 650504, China}
\newcommand{\NJU}{School of Astronomy and Space Science, Nanjing University, Nanjing 210093, China}
\newcommand{\NJUMOE}{Key Laboratory of Modern Astronomy and Astrophysics (Nanjing University), Ministry of Education, Nanjing 210093, China}
\newcommand{\NAOC}{National Astronomical Observatories, Chinese Academy of Sciences, 20A Datun Road, Chaoyang District, Beijing 100101, People’s Republic of China}
\newcommand{\UCAS}{School of Astronomy and Space Science, University of Chinese Academy of Sciences, Beijing 100049, People’s Republic of China}
\newcommand{\USTCA}{Department of Astronomy, University of Science and Technology of China, Hefei 230026, People’s Republic of China}
\newcommand{\FAST}{CAS Key Laboratory of FAST, National Astronomical Observatories, Chinese Academy of Sciences, Beijing 100101, People’s Republic of China}
\newcommand{\HKIAA}{Hong Kong Institute for Astronomy and Astrophysics,
University of Hong Kong, Pokfulam Road, Hong Kong, China}
\newcommand{\UNLV}{The Nevada Center for Astrophysics, University of Nevada, Las Vegas, NV 89154, USA}
\newcommand{\UNLVP}{Department of Physics and Astronomy, University of Nevada, Las Vegas, NV 89154, USA}
\newcommand{\HKU}{Department of Physics, University of Hong Kong, Pokfulam Road, Hong Kong, China}

\begin{document}

\title{A possible periodic RM evolution in the repeating FRB 20220529}

\author{Yi-Fang Liang}
\affiliation{\PMO}
\affiliation{\USTC}

\author{Ye Li$^*$}
\affiliation{\PMO}

\author{Zhen-Fan Tang}
\affiliation{\PMO}
\affiliation{\USTC}

\author{Xuan Yang$^*$}
\affiliation{\PMO}

\author{Song-Bo Zhang}
\affiliation{\PMO}

\author{Yuan-Pei Yang}
\affiliation{\YNU}
\affiliation{\PMO}

\author{Fa-Yin Wang}
\affiliation{\NJU}
\affiliation{\NJUMOE}

\author{Bao Wang}
\affiliation{\PMO}
\affiliation{\USTC}

\author{Di Xiao}
\affiliation{\PMO}

\author{Qing Zhao}
\affiliation{\PMO}
\affiliation{\USTC}

\author{Jun-Jie Wei}
\affiliation{\PMO}
\affiliation{\USTC}

\author{Jin-Jun Geng}
\affiliation{\PMO}

\author{Jia-Rui Niu}
\affiliation{\NAOC}

\author{Jun-Shuo Zhang}
\affiliation{\NAOC}
\affiliation{\UCAS}

\author{Guo Chen}
\affiliation{\PMO}

\author{Min Fang}
\affiliation{\PMO}

\author{Xue-Feng Wu$^*$}
\affiliation{\PMO}
\affiliation{\USTC}

\author{Zi-Gao Dai}
\affiliation{\USTCA}
\affiliation{\USTC}

\author{Wei-Wei Zhu}
\affiliation{\NAOC}

\author{Peng Jiang}
\affiliation{\NAOC}
\affiliation{\FAST}

\author{Bing Zhang}
\affiliation{\HKIAA}
\affiliation{\HKU}
\affiliation{\UNLV}
\affiliation{\UNLVP}

\correspondingauthor{Ye Li, Xuan Yang, Xue-Feng Wu}
\email{yeli@pmo.ac.cn;  yangxuan@pmo.ac.cn; xfwu@pmo.ac.cn}

\begin{abstract}
Fast radio bursts (FRBs) are mysterious millisecond-duration radio transients of extragalactic origin. Some of them repeat, while others apparently do not. Investigations of periodic activity in repeating FRB have been conducted to probe their origins.
While periodicity in the burst rate has been reported, studies of periodicities in other properties, such as dispersion measure (DM) and rotation measure (RM), are sparse.
FRB~20220529 was monitored by the Five-hundred-meter Aperture Spherical radio Telescope (FAST) for nearly three years, providing an opportunity to investigate periodicity in its observed properties.
Here we report a possible period of $\sim 200$ days in the RM evolution, with a significance of {4.1 $\sigma$} estimated via the Lomb-Scargle algorithm and {3.1 $\sigma$} with the phase-folding method. 
Periodicity in the burst rate was also investigated. 
It may indicate that the FRB progenitor is in a binary system, which is consistent with the significant RM increase and prompt recovery of this FRB on a week-timescale. Other scenarios, such as a system with an intermediate-mass black hole, are also explored. 

\end{abstract}

\section{Introduction}

FRBs are extragalactic radio transients with durations on the order of milliseconds (ms) (\citealt{lorimer07,Thornton2013}, see \citealt{zhang2023review} for a review). Many theoretical models have been proposed to explain FRBs (see \citealt{katz16, platts2019} for theoretical reviews\footnote{https://frbtheorycat.org/index.php/Main\_Page}), with most of them invoking neutron stars or other compact objects (e.g., black holes or white dwarfs). To date, about 1000 FRBs have been reported. Among them, around 60 are known to repeat \citep{chime-repeaters,chime2023rfrb}, exhibiting burst counts ranging from 2 to over $10000$ \citep{lidi2021, Xu2022, niuc2022, Zhoudj2022, Zhangyk2023, Zhoudj2025}, others appear non-repeating \citep{spitler16, lidi2021}. 
The evolution of repeating FRB properties, such as burst rate, dispersion measure (DM), and rotation measure (RM), provides invaluable diagnostics of their progenitors and local environments. 
While most of them seem to be irregular, a few repeating FRBs show interesting properties.

Observations of repeating FRBs have revealed periodic activity in a few sources.
The first FRB reported to exhibit periodic activity was FRB~20180916B, which shows a well-established period of 16.35 days \citep{chime180916}.
The active phase has been further reported to be frequency-dependent \citep{pastor2021,Bethapudi2023}. 
In addition, a possible period of approximately 160 days was identified in the active repeater FRB~20121102A \citep{Rajwade2020,Cruces2020, braga2025}.
Recently, a candidate period of 126 days was reported for FRB~20240209A \citep{Arpan2025}. 
These discoveries highlight the diversity in the recurrence timescales and modes of repeating FRBs. 
The presence of periodicity in a few repeaters suggests that at least some FRBs may originate from binary systems \citep{Ioka_zhang2020, Wada2021,ZhangB2025} or undergo precessing motion \citep{levin20}.

If FRBs are from binary systems, periodic variation of other properties, such as RM, is also expected \citep{wangfy2022,Zhang2018,yangyp2023,Lirn2025}. 
Physically, RM quantifies the convolution of electron density and magnetic field along the line of sight, serving as a diagnostic of the magneto-ionic environment surrounding an FRB source. 
For FRB~20180916B, the RM exhibits stochastic fluctuations around $\rm -114~rad~m^{-2}$ before MJD 59300. Then, it nearly linearly changes to $\rm -50~rad~m^{-2}$ from MJD 59300 to 59700, followed by random variations until 60352 \citep{Mckinven2023, bethapudi2025}. 
Notably, this RM evolution lacks periodicity correlated with the source's burst activity cycle. 
Potential explanations include the expansion of a supernova remnant or turbulent plasma fluctuations in the vicinity of the FRB \citep{yangyp2023}. 
Alternatively, fitting the secular RM component with binary models yields an orbital period of $>3$ years \citep{yangyp2023, Zhao2023}, which differs significantly from the 16.35-day activity period.
The RMs of FRB~20121102A have absolute values as high as $\rm \sim 10^5~rad~m^{-2}$. Long-term monitoring has revealed a decreasing RM trend accompanied by some fluctuations \citep{Michilli2018,Hilmarsson2021}. 
Similar to 20180916B, no periodic RM evolution has been detected. 
The RM evolution of FRB~20121102A may be attributed to a supernova remnant or the wind nebula surrounding the FRB progenitor.
Additional RM studies of other repeating FRBs have also provided important insights into the origins of FRBs. 
For instance, irregular variations have been observed in FRB~20201124A \citep{Xu2022}, while a sign reversal in RM has been detected in FRB~20190520B \citep{Anna-Thomas2023}. 
Nevertheless, periodic studies of RM are sparse, with a recent exploration of FRB~20201124A \citep{xujw2025}.

FRB~20220529 is another repeating FRB that may be associated with a binary system. It was first detected by the Canadian Hydrogen Intensity Mapping Experiment (CHIME) on May 29, 2022, and was identified as a repeating FRB originating from a disk galaxy at a redshift of 0.1839.
We monitored it using the Five-hundred-meter Aperture Spherical radio Telescope (FAST) and the Parkes telescope for nearly three years.
It shows an abrupt increase in RM to $\rm \sim 2000\,rad\,m^{-2}$ and prompt recovery of RM within two weeks, against a baseline of $\rm 17\pm101\,rad\,m^{-2}$ for more than 1.5 years, strongly supporting a binary origin \citep{li2025}.
The continuous detection of pulses in FRB~20220529 enables us to examine the periodicity of its properties more efficiently.

In this work, we perform a periodicity search for the RM evolution of FRB~20220529, which spans approximately 1000 days. Using both the Lomb-Scargle Periodogram (LSP) method and the period-folding analysis, we investigate potential periodic modulation in the RM variations of this actively repeating FRB.
The paper is structured as follows.
The observations and data reductions of FRB~20220529 are presented in Section \ref{sec:observe}. 
The periodicity examinations of its RM are presented in Section \ref{sec:periodicity}.
The implications of our results are discussed in Section \ref{sec:discussion}, and we conclude in Section \ref{sec:conclusion}.

\section{Observations and Data Reduction}\label{sec:observe}

FRB 20220529 was discovered by the CHIME telescope, and was reported to repeat in June 2022. It has been monitored with the FAST telescope since June 22, 2022. Besides the four-grid observations on June 22, August 14, and August 17, 2022, as well as an off-beam tracking observation on August 28, 2022, FRB~20220529 was observed in tracking mode with the center beam of the FAST telescope, covering a frequency range of 1000-1500 MHz with 4096 channels. Up to April 1st, 2025, 125 observations totaling 58.4 hours were conducted with FAST, including 52.4 hours of on-source tracking. 
\cite{li2025} reported the observations before September 5th, 2024. Here we summarize the FAST observations from September 5th, 2024 to April 1st, 2025 in Table \ref{tbobs}. 
During this time, 13 observations were conducted, totalling 4.5 hours of exposure.

Data collected from FAST were analyzed using two independent searching pipelines built upon the pulsar/FRB single-pulse searching packages \texttt{PRESTO} \citep{Ransom2001} and \texttt{HEIMDALL} \citep{heimdall} ~\footnote{https://sourceforge.net/projects/heimdall-astro/}, which processed the full-band data. The DM was searched over the range of $200–300$\,pc\,cm$^{-3}$ with a step size of 0.1\,pc\,cm$^{-3}$. Single-pulse candidates with signal-to-noise ratios (SNR) greater than 7 were retained and manually verified.
A total of 1169 bursts were detected during the entire monitoring period, with 1094 occurring when the observations were on source.  
Among them, 1156 bursts were reported in \citealt{li2025}, while the remaining 13 bursts, detected from September 5, 2024 to April 1, 2025, are presented in Table~\ref{tbobs}.
After October 21, 2024, only one burst was detected, on February 22, 2025.

After being dedispersed at the detection DM with the maximum SNR,
the polarization data were calibrated using the \texttt{PSRCHIVE \citep{Hotan2004PSRCHIVE}\footnote{https://psrchive.sourceforge.net/}} software package, with differential gain and phase corrections applied via a noise diode signal injected prior to each observation.
The RM value of each burst was obtained using the \texttt{RMFIT} program by searching for the peak of the linear polarization intensity, $ L = \sqrt{Q^2 + U^2} $, within the range of -4000 to +4000 $\mathrm{rad\,m^{-2}}$ with a step size of $1\,\mathrm{rad\,m^{-2}}$. \texttt{RMFIT} applies Faraday rotation correction for each trial RM and selects the RM that maximizes the linear polarization as the final measurement \citep[see][for more details]{li2025}.

The updated RM values are presented as blue dots in Figure \ref{fig:rm}. 

\input{tbobs}

\begin{figure*}[!htbp]
    \centering
    \includegraphics[width=0.85\linewidth]{rmflare_basic20250520.pdf}
    \caption{Temporal Evolution of RM, DM, and Burst Rate in FRB~20220529. (A) FRB Daily Burst Rate. Grey vertical dashed lines indicate observation days (bursts detected and undetected). The y-axis of the plot is logarithmic. (B) Dispersion measure of bursts. The blue points represent the detected DM values with their uncertainties. (C) Rotation measure of bursts. The filled blue region delineates the RM variation range excluding the RM flare epoch, with the y-axis range scaled to compress the RM flare region and enhance the visibility of the primary dataset. The red points correspond to the daily-binned RM values. The orange lines mark the phases corresponding to the peak of the 200-day period, approximately at phase 0.25.}
    \label{fig:rm}
\end{figure*}

\section{Periodicity Examinations}\label{sec:periodicity}

In order to explore the possible periodicity of FRB~20220529, we examine the periodicity using the RM series described in Section \ref{sec:observe}.
As reported in \cite{li2025}, FRB~20220529 experienced an abrupt change and prompt recovery of RM at the end of 2023, reaching $\sim 20~\sigma$ of the RM baseline (from $\rm -300~rad~m^{-2}$ to $\rm +300~rad~m^{-2}$). We therefore exclude those pulses within the ``RM flare'' epoch and with $\rm RM > 300~rad~m^{-2}$ in our periodicity studies, specifically $\rm 60290 <MJD< 60310$ as they are inconsistent with any periodic pattern. 
Due to the fluctuations of the magnetized plasma, 
the RMs exhibit significant daily variations. We thus binned the RMs within one day by 
 $\mathrm{RM}= \mathrm{\overline{RM}(MJD)}$.

The resulting daily-binned RMs are plotted as pink points in Figure \ref{fig:rm}.
The periodicity examinations with the LSP and phase-folding analysis are presented in Section \ref{ls} and \ref{chi2}, respectively.

\subsection{Lomb-Scargle Periodogram} \label{ls}

The Lomb-Scargle periodogram \citep{lomb1976, scargle1982} is a powerful method for detecting periodic signals in unevenly sampled time-series data. It estimates spectral power over a range of trial periods, identifying periodicities by evaluating their statistical significance. The spectral power is defined as
\begin{equation}
\begin{split}
\rm P_{\mathrm{LS}}(\omega)& =\frac{1}{2}\Bigg{\{}\frac{[\sum_{i}\mathrm{RM_{i}}\cos\omega(t_{i}-\tau\big)]^{2}}{\sum_{i}\cos^{2}\omega\big{(}t_{i}-\tau\big{)}}
\\ & \quad+\frac{\left[\sum_{i}\mathrm{RM_{i}}\sin\omega\big{(}t_{i}-\tau\big{)}\right]^{2}}{\sum_{i}\sin^{2}\omega\big{(}t_{i}-\tau\big{)}}\Bigg{\}},
\end{split}
\end{equation}
where $\omega$ is the angular frequency, $\mathrm{RM_{i}}$ is the observed RM value of the $i$-th time point, $t_{\rm i}$ is the time of the $i$-th data point, and $\tau$ is specified for each $\omega$ to ensure the periodogram's invariance under time translations:
\begin{equation}
    \tau=\frac{1}{2\omega}\tan^{-1}\left(\frac{\sum_{i}\sin2\omega t_{i}}{\sum_{i}\cos2\omega t_{i}}\right),
\end{equation}
where $i$ ranges from 1 to $N$ in the summation. $N$ is the total number of RM points. By normalizing the spectral power, the algorithm quantifies the likelihood of periodicity and determines the most prominent candidate. 

This method is particularly well-suited for astronomical time-domain studies, where observations are often sparse and irregularly sampled. Unlike the classical Fourier transform, the LSP effectively handles data gaps and provides unbiased estimates of periodic signals \citep{VanderPlas2018}. We implement the LSP using the 
\texttt{LombScargle} class from the \texttt{astropy}\citep{astropy:2022} package\footnote{https://docs.astropy.org/en/stable/timeseries/lombscargle.html}. 
We compute the false alarm probability (FAP), which quantifies the likelihood that an observed peak in the power spectrum arises purely from random noise under the null hypothesis. The power threshold corresponding to a given FAP is determined using the \texttt{false\_alarm\_probability} method of the \texttt{LombScargle} class.

The results of FRB~20220529 are shown in upper left panel of Figure \ref{fig:ls_pf}. Using a FAP threshold of $10^{-4}$, we identified the most significant period as $\rm P = 199 ^{+10}_{-10}$ days. The uncertainty was estimated from 1000 Monte Carlo simulations, where each RM measurement was resampled based on its observational uncertainty. The distribution of the best-fit periods from these simulations yields the quoted confidence interval.

The analytic FAP relies on idealized assumptions, such as Gaussian noise and approximately uniform sampling, which are not satisfied in our data. To obtain a more reliable significance under realistic conditions, we employed a Bootstrap resampling procedure. Specifically, we preserved the observing times while randomizing the observed RMs, and then calculated the maximum LS power in each realization. Based on $10^6$ simulations, the null probability was estimated as the fraction of trials with maximum powers larger than the observed one.
The resulting probability is $2\times10^{-5}$, indicating that the likelihood of obtaining a signal of equal or greater power by random chance is very small, although a purely noise origin cannot be fully excluded. Converting this probability into an equivalent Gaussian significance yields 4.1~$\sigma$. 
The distribution of simulated maximum powers is shown in the upper right panel of Figure \ref{fig:ls_pf}.

\begin{figure*}[ht]
    \centering
    \includegraphics[width=0.45\linewidth]{ls1_1016.pdf}
    \includegraphics[width=0.45\linewidth]{ls_simual_1016.pdf}
    \includegraphics[width=0.45\textwidth]{chi2_1030.pdf}
    \includegraphics[width=0.45\textwidth]{simul_10_chi2.pdf}
    \caption{Results of Periodicity examinations. Upper Left: LSP of RM variations for FRB~20220529, derived from FAST observations. 
    The red dashed line marks the most significant peak at 199 \text{days}. The x-axis of the plot is logarithmic. 
    Upper right: Simulation results of the LSP. The red dashed line represents the maximum power obtained with the observed RM. In the simulation, RM values were randomly selected and assigned to the TOA before applying the LSP. $10^6$ simulations were performed. The probability of obtaining a maximum power greater than or equal to the observed maximum power is {$2\times10^{-5}$}. 
    Lower Left: Results of phase folding analysis. The red dashed line indicates the period corresponding to the maximum $\chi^2$ value, which is {204} days. The x-axis of the plot is logarithmic. Lower Right : Simulation results of phase folding analysis. $10^6$ simulations were performed.
    The probability of obtaining a maximum $\chi^2$ greater than or equal to the observed maximum $\chi^2$ is {$1.0\times10^{-3}$}.
 }
    \label{fig:ls_pf}
\end{figure*}

\subsection{Phase Folding method} \label{chi2}

Although the LSP can handle unevenly sampled data, it implicitly assumes a sinusoidal signal. Since the RM evolution is not necessarily sinusoidal, $\chi^2$ phase-folded analysis is used as a complementary method. This method evaluates how strongly the folded RM values deviate from a uniform distribution, without assuming any specific functional form. If the RMs are purely random, their folded distribution is expected to be statistically uniform, leading to a small $\chi^2$ on average; if they are periodic and the correct trial period $P$ is used, the folded RMs deviate from uniformity, resulting in a larger $\chi^2$.

The procedure is as follows. First, we assume a trial period $P_0$ and adopt the time of the first RM measurement as the reference epoch $t_0$. The times of RMs $t_i$ are then folded into phases:
\begin{equation}
    \phi_i = \frac{(t_i - t_0)\bmod {P_0}}{P_0},
\end{equation}
where $t_i$ is the observing time of the $i$-th RM measurement. Second, the folded RM values are grouped into $n$ phase bins. Third, we compute the $\chi^2$ statistic as
\begin{equation}
    \chi^{2}=\sum_{j=1}^{n}\frac{(\mu_j-\bar{\mu})^{2}}{\sigma_j^2},
\label{eq:chi2}
\end{equation}
where $\mu_j$ is the mean of the RM values in the $j$-th bin, and $\bar{\mu}$ is the global mean of all RM measurements (equivalent to the mean of RMs). The uncertainty $\sigma_j$ for each bin incorporates both the median measurement error of the constituent points and the intrinsic scatter of the RM values within that bin.
To ensure statistical robustness, we implement two criteria. First, a lower limit is applied to $\sigma_j$ in Equation~\ref{eq:chi2}, set to the median of the individual RM measurement errors across the entire dataset. This prevents the $\chi^2$ from being dominated by bins with a spuriously small standard deviation. Second, each phase bin is required to contain a minimum of four data points.

In practice, a large $\chi^2$ indicates that the trial period $P_0$ is a plausible candidate for the intrinsic period, whereas a small $\chi^2$ suggests that the period is unlikely. By systematically scanning $P_0$ across a range of trial periods and evaluating the statistical significance, we can identify the most probable intrinsic period, if it exists.

In our work, we bin the RM values of daily bursts into a single value for each day. We then test periods within the range of 5 to 500 days, dividing each trial period into {10} bins. This approach leads to the identification of a best-fit period of  {$204^{+8}_{-11}$ days}. The uncertainty of the period is also derived by Monte Carlo simulations. The corresponding $\chi^2$ distribution is shown in the lower left panel in Figure \ref{fig:ls_pf}. Finally, we also perform an MCMC simulation to determine the SNR of the identified period. For each simulation, we compute the $\chi^2$ value, and record the maximum $\chi^2$. After conducting $10^6$ simulations, we find that the probability of obtaining a maximum $\chi^2$ matching the observed one is about {$1.0\times10^{-3}$}, corresponding to ${\rm SNR}=3.1~\sigma$. The result of the simulation is shown in the lower right panel of Figure \ref{fig:ls_pf}.

To examine whether the period is real, we fold the RM evolution with a 200-day period and present the folded RM distribution as a function of phase in Figure \ref{fig:fold_200}. 
To study the pattern, we fited the folded RM with a sinusoidal function. The best-fitting function is 
\begin{equation}
\mathrm{RM}=\mathrm{RM_0}+\mathrm{RM_A}\sin\left( 2\pi \frac{t-\mathrm{MJD_0}}{\rm P}\right).
\label{eq:sinfit}
\end{equation}

with the best-fitting parameters
$\mathrm{MJD_0=59828}$, $\mathrm{RM_A=76.97~rad~m^{-2}}$ and $\mathrm{RM_0=83.53~rad~m^{-2}}$, equivalent to $\bar\mu$ in Equation \ref{eq:chi2}.
We label the peaks as vertical yellow dashed lines in Figure \ref{fig:rm}. It can be seen that the predicted peaks generally match the four peaks in the RM evolution. However, we note that the RM evolution is not necessarily sinusoidal; Equation~\ref{eq:sinfit} only provides a rough guide to the peak locations.

\begin{figure}[!htbp]
    \centering
    \includegraphics[width=0.96\linewidth]{folded_fit_10.pdf}
    \caption{Phase-folded RM variations with a 200-day period. The blue points represent the observed RM values and their uncertainties. The red points show the averaged RM values within each phase bin. The orange line depicts the fit result to a sine function.}
    \label{fig:fold_200}
\end{figure}

\section{Discussion} \label{sec:discussion}

\begin{figure}[!htb]
    \centering
    \includegraphics[width=0.96\linewidth]{chi2_simul_nbin_1030.png}
    \caption{Significance for different bin sizes and periods in phase-folding periodicity search. The x-axis represents the bin size, and the y-axis shows the period. The color intensity indicates the corresponding significance, which reflects the significance of the periodicity detection for each trial. It is evident that periods around $\sim$ 200 days are relatively significant for reasonable bin sizes.}
    \label{fig:binsize}
\end{figure}

\subsection{Effect of the ``RM flare''}

In our investigations, we excluded the data with $\rm |RM|>300~rad~m^{-2}$ within the ``RM flare'' epoch, specifically $\rm 60290 < MJD < 60310$. We choose $300~{\rm rad~m^{-2}}$ as the upper limit since it is generally 3 $\sigma$ of the RM standard deviation and includes nearly all other data. Still, we tested the results with all the data, including the ``RM flare''. The LSP still presents the $\sim 200$-day periodicity as the most prominent one, with a significance of {$3.8~\sigma$}. In contrast, the phase-folding method is more strongly affected by the inclusion of the RM flare. It reveals that the LSP could deal with outliers much better than the phase-folding method. Given that the RM flare is clearly not part of the underlying periodic RM evolution, it is more appropriate to exclude this segment from the analysis.

\subsection{Effect of background subtraction}

In some analyses, background subtraction is performed prior to searching for periodicity, motivated by the possibility that long-term variations could mask periodic signals. For example, if an FRB has a companion star undergoing episodic mass ejection, the combination of mass ejection with orbital motion could produce periodic RM variations on top of an evolving background.

To explore this effect, we fitted a linear trend to the observed RM data and subtracted it before applying the Lomb–Scargle Periodogram. The best-fit background is given by $\mathrm{RM_b}=(21520-0.3586~\mathrm{MJD})\ \mathrm{rad~m}^{-2}$. After subtraction, the significance of the 199-day signal increased to $5.2\sigma$. A second polygon background subtraction would increase it to $5.4~\sigma$.

These tests show that background subtraction can affect the statistical significance of the signal. However, without a clear physical model for the background evolution, the choice of subtraction scheme becomes arbitrary and may bias the results. We therefore present these tests for completeness but do not adopt them in our main conclusions.

\subsection{Limitation of the Lomb-Scargle periodogram}

Despite the capability of the LSP to handle unevenly sampled data, it also has drawbacks. Specifically, it has the tendency to generate spurious peaks and obscure genuine periodic signals, thereby compromising the accuracy of periodicity analysis. 

Firstly, the LSP is based on the assumption that the underlying signal is sinusoidal. Consequently, its detection capability for non-sinusoidal periodic signals or those containing small-scale structures is significantly reduced.
As illustrated in Figure \ref{fig:rm}, the RM evolution does not follow a standard sinusoidal pattern and exhibits irregular variations, such as the peak near MJD 60000.
These deviations from the sinusoidal form give rise to multiple peaks in the LSP. For instance, a prominent secondary peak is observed at approximately the 130-day period.
According to Monte Carlo simulations, this peak has a statistical significance of 3.0 $\sigma$.
Furthermore, the RM evolution around MJD 59800 resembles a compressed portion of a sine wave and displays numerous minor peaks. These minor peaks in the RM data generate multiple small peaks in the LSP, such as the peak around the 20-day period. The cumulative effect of these peaks in the RM curve leads to fluctuations in the LSP.

Secondly, the LSP is sensitive to irregular sampling. 
As illustrated in Figure \ref{fig:rm}, the sampling rate before MJD 60050 is significantly higher than that after it. This is primarily due to the increased FRB activity. As reported in \cite{li2025}, FRB~20220529 exhibited two active bursting epochs around MJD 59800 and MJD 60000. Moreover, the community typically schedules intensive observations during epochs of high burst rates, further enhancing the unevenness of the sampling. 
An unbinned analysis tends to downplay the contributions of data points collected during epochs of low FRB activity, especially after MJD 60050.
In addition, RM is inherently subject to strong fluctuations. 
Since periodic studies require capturing the long-term trends, we binned RM values on a daily basis before performing the periodicity analysis. 
In order to assess the impact of the binning, we also conducted a periodicity analysis using unbinned RM data. The LSP reveals that the peak near $198$ days persists, while the second peak in the upper left panel of Figure \ref{fig:ls_pf} becomes more pronounced. This is a result of uneven sampling. 
It is evident that daily binning of RM data provides a clearer depiction of periodic behavior over extended timescales. 
Although an evenly sampled RM dataset could ideally improve the periodicity analysis, achieving this is challenging. Even with regularly scheduled observations, it is inherently constrained by the irregular nature of FRB activity.

\subsection{Limitation of the phase folding method}

When using the phase folding method to search for periodicity, we initially adopted a bin size of 10 to calculate the $\chi^2$. Here we tested the impact of different bin sizes on the results. The significance for different bin size and periods are presented in Figure \ref{fig:binsize}. It is shown that with bin sizes from 5 to 16, the periods around 200 have significance larger than 3 $\sigma$. The results indicate that within a reasonable range, the bin size does not significantly affect the periodicity detection. It is important to note that the significance (in terms of sigma) in Figure \ref{fig:binsize} represents the relative strength of the trial periods, rather than the actual SNR of the detected periodicity. The final result presented in Figure \ref{fig:ls_pf}.

\subsection{Effect of sampling}

In order to investigate whether the periodic signals detected in the RM time series are dominated by the sampling process, we further examined the periodicity of the burst rate of the pulses. The FAP derived from the LSP of the burst rate and the RM evolution are shown as the blue and orange lines, respectively, in Figure~\ref{fig:fap}. 
To make a direct comparison between the results of the burst rate and RM, we plot the FAP instead of power here. 
It turns out that there is also a peak at approximately 200 days in the LSP of the burst rate, with $\rm FAP = 0.17$ (equivalent to $1\sigma$), hinting at a tentative $\sim 200$-day periodicity. However, this peak is not statistically significant. In contrast, the RM dataset shows a much stronger signal at $P_0 = 199$ days, with $\rm FAP = 5 \times 10^{-5}$ (equivalent to $3.9~\sigma$), indicating significantly higher statistical confidence than that for the burst rate periodicity. This clearly demonstrates that the periodicity in RM is much more robust than that in the burst rate. Note that this FAP estimate differs slightly from the value obtained via our bootstrap simulation.
Thus, the periodic evolution of the RM is neither dominated by the sampling process nor by the burst rate. 
The periodic evolution of the RM is more likely inherent to the system, if it is not merely coincidental. 
Assuming that the RM and FRB rate share the same 200-day period, we examine the phase-folded RM variation (Figure~\ref{fig:fold_200}).  
The active bursting epochs, occurring at MJDs $\sim59800$ and $\sim60000$, correspond to a phase of approximately 0.86, where the RM appears smaller. However, this trend is not statistically significant, as only two bursting epochs were observed.

\begin{figure}[!htb]
    \centering
    \includegraphics[width=0.96\linewidth]{fap_rate_rm_1016.pdf}
    \caption{False Alarm Probability versus trial period in LSP. Orange points show FAP values for each trial period in the RM periodicity search, while blue points denote results from the Burst Rate periodicity analysis. Gray dashed lines indicate statistical significance thresholds corresponding to $3\sigma$ and $5\sigma$ confidence levels. The y-axis is inverted, with FAP values decreasing from bottom to top.}
    \label{fig:fap}
\end{figure}

\subsection{Physical implications}

The possible 200-day periodic RM evolution revealed in our analysis carries significant implications for understanding the origin of FRBs.
As a convolution of the electron density and magnetic field in the line of sight, the observed RM of an FRB from a redshift $z$ is attributed to various plasma components, 
$\rm  RM_{obs} = RM_{ion} +RM_{MW} +RM_{IGM} + RM_{host}/(1+z)^2 + RM_{loc}/(1+z)^2$. 
Here, $\rm RM_{ion}$ denotes the contribution from the Earth's ionosphere, typically on the order of $\rm 0.1-1~rad~m^{-2}$ \citep{Mevius2018}. 
$\rm RM_{MW}$ represents the contribution from the interstellar medium within the Milky Way. 
In the case of FRB~20220529, this value is approximately $\rm -35~rad~m^{-2}$ \citep{Oppermann2015}. 
$\rm RM_{IGM}$ accounts for the contribution from the intergalactic medium, which is typically $\rm < 10~rad~m^{-2}$ \citep{Akahori2016}. 
$\rm RM_{host}$ is the contribution from the interstellar medium within the FRB's host galaxy, potentially comparable in magnitude to that of the Milky Way. Lastly, $\rm RM_{loc}$ signifies the contribution from the local plasma in the immediate vicinity of the FRB source. Therefore, an observed RM with an absolute value $\rm > 10^2~rad~m^{-2}$ is expected to be mainly contributed by the local plasma $\rm RM_{loc}$. In addition, the variation timescales of interstellar and intergalactic medium are expected to be long. The possible RM variation observed here is expected to be a result of the local magneto-ionic environment.  

It is natural to consider that the periodic evolution of the RM is generated by the orbital configuration of a binary system.
If the FRB progenitor resides in a binary system, 
the stellar winds from a massive/giant star companion would contribute to the RM \citep{wangfy2022, yangyp2023}. 
Thus, the orbital motion or the dynamic evolution of the wind would result in RM variations, 
and the RM variation due to the orbital motion should have the same period as the orbital period.
The binary system model is further supported by the ``RM flare'' reported in \cite{li2025}. 
This ``RM flare'' could be explained by two possible scenarios: (1) a coronal mass ejection (CME) from the companion or a disk, or (2)
an orbital configuration with a period $>1000$ days and an eccentricity $>0.9$. If the $\sim$200-day periodicity is indeed intrinsic and caused by orbital motion, then the orbital scenario cannot easily explain the ``RM flare'', and the CME origin becomes the more plausible explanation for the “RM flare.”
The binary model is also consistent with the peak at approximately $200$ days in the LSP of the burst rate, which indicates tentative periodic activity that aligns with the periodic RM evolution. 

In a binary system comprising a companion star with mass $M_c$, radius $R_c$, surface magnetic field $B_c$, orbital period $P$ and separation $a$, the RM contribution due to the stellar wind can be estimated as \citep{yangyp2023}
$$\mathrm{RM}_w=\frac{e^3}{2\pi m^2_ec^4} \int B n_e dl\sim \frac{e^3B n_w r}{2\pi m^2_ec^4},$$
where $m_e$ is the electron mass, $c$ is the speed of light, $e$ is the electron charge. The magnetic field $B = B_c(r/R_c)^{-\beta_\mathrm{B}}$, with $\beta_\mathrm{B}=1$ for toroidal field and $\beta_\mathrm{B}=2$ for radial field. The electron density in the stellar wind at a distance $r$ is given by 
$$n_w=\frac{\dot{M}}{4\pi \mu_m m_p v_w r^2},$$ 
where $\dot{M}$ is the mass loss rate, $\mu_m=1.2$ is the mean molecular weight, $m_p$ is the proton mass, and $v_w$ is the speed of the wind, which could be estimated as the escape speed $v_w=(\frac{2GM_c}{R_c})^{1/2}$. 
As discussed in \citep{yangyp2023}, the $\rm RM_w$ is dominated by the wind around the binary separation $r \sim a$, where Kepler's Law gives $a=(GM_\mathrm{tot}(\frac{P}{2\pi})^2)^{1/3}$. Thus, we have the relation among the magnetic field $B$, companion mass $M_c$, the mass loss rate $\dot{M}$, the RM contribution of the wind $\mathrm{RM}_w$, and the period $P$
\begin{equation}
\label{eq:theory}
\mathrm{RM}_w (\frac{P}{2\pi})^\frac{2}{3}=\frac{e^3B}{2\pi m^2_ec^4}\frac{\dot{M}}{4\pi \mu_m m_p}(\frac{R_c}{2GM_c})^\frac{1}{2}(\frac{1}{GM_\mathrm{tot}})^\frac{1}{3}
\end{equation}
Assuming the orbital period as the observed period $P=200$ days and the contribution of the wind as the observed $\sigma_\mathrm{RM}=100$ rad m$^{-2}$, we estimate the surface magnetic field $B_c$ for different assumed values of $M_c$ and $\dot{M}$,  as presented in Figure \ref{fig:theory}. The stellar mass and mass loss rate of main-sequence stars are overplotted as dots for comparison \citep{Snow1981, Mokiem2007, Cranmer2011, Wood2021}. \cite{Cranmer2011} provides the stellar mass information, we thus plot their main-sequence stars as black dots and evolved stars as gray circles. For objects from \cite{Snow1981} and \cite{Mokiem2007}, the stellar masses are estimated with $M_c \propto T_\mathrm{eff}^{0.55}$ and a solar mass of 5772 K is adopted. For objects from \cite{Wood2021}, the stellar masses are estimated with $M_c \propto R_c^{0.7}$. For binaries, the mass loss rates are counted separately if values for both stars are available. The mass loss rates are used as the values of the more massive star if only one is available. It is shown that B stars with low magnetic fields, M stars with high magnetic fields or median mass stars with median magnetic fields would be consistent with the period and RM we detected.

\begin{figure*}[ht]
    \centering
    \includegraphics[width=0.49\linewidth]{theory1.png}
    \includegraphics[width=0.49\linewidth]{theory2.png}
    \caption{The surface magnetic field $B_c$ for different assumed values of the companion mass $M_c$ and mass loss rate $\dot{M}$. The observed period $P=200$ days is assumed to be the orbital period, and $\sigma_\mathrm{RM}=100$ rad m$^{-2}$ is taken as the RM contribution of the stellar wind. The observed stellar mass and mass loss rate of main-sequence stars as well as evolved stars are overplotted as black dots and gray circles for comparison. The left and right panels show the cases for a toroidal field ($\beta_B=1$) and a radial field ($\beta_B=2$), respectively.  
 }
    \label{fig:theory}
\end{figure*}

Also, the periodic RM variation could also arise from magnetized outflows or a disk associated with a massive black hole \citep{Zhang2018, yangyp2023}. If the orbit of the FRB source is elliptical, a large RM variation would occur near the periastron and remain nearly constant far away from the periastron. A periodic evolution of RM variation could result from the orbital motion of the FRB sources. 
However, FRB~20220529 is not located at the center of its host galaxy. The distance between FRB~20220529 and the center of the host galaxy exceeds the half-light radius of the host. Moreover, outflows from a massive black hole typically induce an RM of $\rm \gtrsim 10^4~rad~m^{-2}$, whereas the RM amplitude observed for FRB~20220529 in this study is around $\rm 100~rad~m^{-2}$. According to Equation (50) of \cite{yangyp2023}, a black hole with a mass of $10^3~M_{\odot}$ would produce an RM of $\rm \sim 100~rad~m^{-2}$, indicating that an intermediate mass black hole may also account for the periodic RM evolution here. However, since the FRB progenitor is widely believed to be a neutron star, an intermediate-mass black hole scenario would imply a black hole-neutron star system. Such a scenario would necessitate additional components to explain the black hole's outflow or disk, making the model somewhat complex.

At last, if a Faraday screen with an inhomogeneous medium is present near the FRB source, the relative motion between the FRB source and the screen would induce random RM variations. These variations could be quantified using the structure function. \cite{li2025} reveals that the structure function for the RM of FRB~20220529 is consistent with other repeating FRBs, although slightly shallower.  
Given significance levels of $4.1~\sigma$ and $3.1~\sigma$ for the 200--day periodicity, based on the LSP and phase-folding analysis, respectively, 
we cannot rule out the possibility that the observed signal arises from random fluctuations. Longer-term monitoring is still required to confirm or refute the tentative periodicity identified here. 
Moreover, although precession of the neutron star’s spin or magnetic axis may induce a periodic modulation in burst rate, they do not typically result in periodic RM evolution because the medium along the line of sight does not change much. We thus do not discuss precession in detail here.

\section{Conclusion} \label{sec:conclusion}

In this paper, we present an investigation of the periodicity of FRB~20220529 using three years of monitoring data from FAST telescope. 
The RM evolution exhibits a potential periodic modulation with a period of $\rm P = 199 ^{+10}_{-10}$ days (LSP), $\rm P = 204 ^{+8}_{-11}$ days (Phase Folding). The statistical significance is 4.1 $\sigma$ determined by the LSP and 3.1 $\sigma$ from phase-folding analysis. While the discrepancy between the two methods may arise from algorithmic sensitivity to non-sinusoidal signals or uneven sampling, their mutual consistency 
supports the case for periodicity. An examination of burst rate variability also suggests a similar timing feature, though much less significant.

We discuss three plausible scenarios:
(1) Binary Orbital Motion: A binary system with a massive/giant star companion would induce RM periodicity through orbital motion. However, current data insufficiently constrain the companion's mass or orbital parameters. 
(2) Massive Black hole: An intermediate-mass black hole with outflows or an accretion disk would induce periodic RM evolution similar to that identified here. However, this scenario requires additional components to produce the outflow/disk, making the system more complex.
(3) Turbulence: Variations in an inhomogeneous medium along the line of sight would produce random RM variations. Although this scenario is not statistically preferred, it can not be fully ruled out. 
The possible periodic RM evolution identified here requires extended monitoring to validate it.

\section*{Acknowledgements}
We thank the anonymous referee for helpful suggestions and comments.
YL thanks Qiang Yuan for helpful discussion. YFL thanks Tiancong Wang for helpful discussion.
This work is partially supported by the Natural Science Foundation of China (Grant Nos. 12321003, 12041306, 12103089,12393813), the National Key Research and Development Program of China (2022SKA0130100), National Key R\&D Program of China(2024YFA1611704), the Natural Science Foundation of Jiangsu Province (Grant No. BK20211000), International Partnership Program of Chinese Academy of Sciences for Grand Challenges (114332KYSB20210018), the CAS Project for Young Scientists in Basic Research (Grant No. YSBR-063), the CAS Organizational Scientific Research Platform for National Major Scientific and Technological Infrastructure: Cosmic Transients with FAST.
X.Y. is supported by the Postdoctoral Fellowship Program of CPSF (grant No. GZC20252100), China Postdoctoral Science Foundation (grant No. 2025M773201), and Jiangsu Funding Program for Excellent Postdoctoral Talent.
Y.P.Y. is supported by the National Natural Science Foundation of China (grant No.12473047), the National Key Research and Development Program of China (2024YFA1611603) and the National SKA Program of China (2022SKA0130100). 
\software{Astropy~\citep{astropy:2013, astropy:2018, astropy:2022}, PRESTO~\citep{2001PhDT.......123R, 2002AJ....124.1788R,2011ascl.soft07017R}, HEIMDALL~\citep[][\url{ https://sourceforge.net/projects/heimdall-astro/}]{heimdall}},PSRCHIVE~\citep[][\url{https://psrchive.sourceforge.net/ }]{psrchive}



\input{ms.bbl}
\end{document}

%% file: tbobs.tex
\begin{table*}[!bht] 
\begin{center}
\caption{FAST observations and the bursts detected after September 5th, 2024}
\label{tbobs}
\begin{tabular}{ccccccc}
\hline \hline
Date & Start MJD & Duration & $\rm N_{\rm FRB}$ & MJD$^{\rm a}$ & RM & DM\\
& & (minutes) & & & ($\rm rad~m^{-2}$)& ($\rm pc~cm^{-3}$) \\\hline 
2024-10-04 & 60587.75208 & 20 & 9 & 60587.7602022 &  $-129\pm20$ & $248.08\pm10.12$ \\
           &             &    &   & 60587.7618459 & - & - \\
           &             &    &   & 60587.7618463 & - \\
           &             &    &   & 60587.7629780 & $-142 \pm 5$ & $247.03\pm0.52$ \\
           &             &    &   & 60587.7669466 & $-146 \pm 3$ & $246.74\pm0.48$ \\
           &             &    &   & 60587.7675164 & $-137.0 \pm 0.1$ & $244.35\pm0.09$ \\
           &             &    &   & 60587.7676134 & $-121 \pm 6$ & $248.40\pm0.59$ \\
           &             &    &   & 60587.7705990 & - \\
           &             &    &   & 60587.7713320 & $-128 \pm 31$ & $250.93\pm0.98$ \\
2024-10-17 & 60600.72292 & 20 & 1 & 60600.7403185 & $-292\pm23$ & $243.84\pm0.27$ \\
2024-11-01 & 60615.58611 & 20 & 2 & 60615.5942126 & $-308 \pm 95$ & $243.89\pm0.54$ \\
           &             &    &   & 60615.5958415 & $-261 \pm 34$ & $243.91\pm0.67$ \\
2024-11-21 & 60635.57153 & 20 & 0 & - & - \\
2024-12-03 & 60647.54167 & 20 & 0 & - & - \\
2024-12-19 & 60663.52986 & 20 & 0 & - & - \\
2025-01-01 & 60676.40625 & 20 & 0 & - & - \\
2025-01-05 & 60680.50972 & 30 & 0 & - & - \\
2025-01-29 & 60704.28611 & 20 & 0 & - & - \\
2025-02-10 & 60716.35486 & 20 & 0 & - & - \\
2025-02-22 & 60728.27083 & 20 & 1 & 60728.2761725 & $-105 \pm 25$ & $241.31$$^{\rm b}$\\
2025-03-09 & 60743.23472 & 20 & 0 & - & - \\
2025-03-22 & 60756.20903 & 20 & 0 & - & - \\
\hline
\end{tabular}
\end{center}
\begin{description}
\item[a] MJDs are in barycentric dynamical time(TDB) and are referenced to infinite frequency.
\item[b] The method we used to estimate the structure-maximizing DM is more accurate for bursts with high S/N and narrow features, we only provide the DM determined by maximizing the S/N of the integrated pulse profile.
\end{description}

\end{table*}